\documentclass[journal]{IEEEtran}

\usepackage{cite}

\ifCLASSINFOpdf
  \usepackage[pdftex]{graphicx}
  \usepackage{color}
  \usepackage[table]{xcolor} 
  \DeclareGraphicsExtensions{.pdf,.jpeg,.png}
\else
  \usepackage[dvips]{graphicx}
  \DeclareGraphicsExtensions{.eps}
\fi
\usepackage{epstopdf}
\usepackage[cmex10]{amsmath}
\usepackage{amsmath}
\usepackage{ dsfont }
\usepackage{mathtools}
\usepackage{breqn}
\usepackage{amssymb}
\usepackage{subfig}
\usepackage{flushend}
\usepackage{tabularx}
\usepackage[font=small]{caption}
\usepackage{booktabs}
\usepackage{glossaries}
\usepackage{amsthm}
\usepackage[ruled]{algorithm2e}
\usepackage[]{algorithm2e}

\definecolor{maroon}{cmyk}{0,0.87,0.68,0.32}

\DeclareMathOperator*{\argmin}{arg\,min}

\begin{document}
%
\title{\textsf{X-FDR}: A Cross-Layer Routing Protocol for \\Multi-hop Full-Duplex Wireless Networks}
%
%
%

\author{M. Omar~Al-Kadri, 
	Adnan~Aijaz, 
	and~Arumugam~Nallanathan 
	\thanks{M. Al-Kadri is with the School of Computing Science and Digital Media, Robert Gordon University, Aberdeen, UK.
	
		A. Aijaz is with the Telecommunications Research Laboratory, Toshiba Research Europe Ltd., Bristol, UK.
		
		A. Nallanathan  is with School of Electronic Engineering and Computer Science, Queen Mary University of London, 	London, UK.
		
		Contact  e-mail: o.alkadri@rgu.ac.uk  }}

%
%

\markboth{IEEE Wireless Communications - Accepted for Publication}%
{Shell \MakeLowercase{\textit{et al.}}: Bare Demo of IEEEtran.cls for Journals}


%



\maketitle

\begin{abstract}
\textcolor{black}{The recent developments in self-interference (SI) cancellation techniques have led to the practical realization of full-duplex (FD) radios that can perform simultaneous transmission and reception. FD technology is attractive for various legacy communications standards. In this paper, after discussing the opportunities of FD technology at the network layer, we present a cross-layer aided  routing protocol, termed as \textsf{X-FDR}, for multi-hop FD wireless networks. \textsf{X-FDR} exploits a Physical (PHY) layer model capturing imperfection of SI cancellation. At the medium access control (MAC) layer, \textsf{X-FDR} adopts an optimized MAC protocol which implements a power control mechanism without creating the hidden terminal problem.  \textsf{X-FDR} exploits the unique characteristics of FD technology at the network layer to construct energy-efficient and low end-to-end latency routes in the network. Performance evaluation demonstrates the effectiveness of \textsf{X-FDR} in achieving the gains of FD at higher layers of the protocol stack. }


\end{abstract}


%
\IEEEpeerreviewmaketitle

\section{Introduction}

\IEEEPARstart{R}{ecent} \textcolor{black}{advances in self-interference (SI) cancellation techniques have made in-band full-duplex (FD) \cite{AnalogSIC2,bharadia2013full} operation feasible for wireless communications. FD-capable nodes can perform simultaneous transmission and reception on same resources in time and frequency domains. FD technology not only offers the potential of (theoretically) doubling the capacity and the spectrum utilization but also assists in solving some of the key problems in half-duplex (HD) networks, such as the hidden node issues, loss of throughput due to high congestion rates, and large end-to-end delays \cite{AnalogSIC2}. Existing efforts towards FD communications have mainly investigated Physical (PHY) layer aspects; however, solutions for medium access control (MAC) and highers layers have also started to emerge \cite{FDsurvey}. In order to reap the maximum benefits of FD technology, optimizations are required at different layers of the protocol stack.  }

On the other hand, energy saving in distributed wireless networks is of significant importance due to the limited battery supply of each node. Nodes in the network continuously participate in route construction, and act as relays for neighboring nodes. In addition to continuous variation in channel conditions, this leads to a large amount of control messages being exchanged across the network, which potentially entails high energy consumption. Therefore, energy-efficiency in distributed wireless networks is an important issue. \textcolor{black}{Moreover, with the introduction of FD, the issue of energy efficiency  becomes critical owing to additional hardware and processing capabilities of nodes. }

\textcolor{black}{Research on routing protocols for FD wireless networks is still in infancy. In \cite{distributedFD}, Fang \emph{et al.} have proposed cross-layer optimization for opportunistic multi-path routing in FD wireless networks. The route selection problem has been solved under various resource competitions and node constraints. However, the proposed framework assumes perfect SI cancellation. Kato and Bandai \cite{directionalFD} have proposed an on-demand detour routing protocol for directional FD wireless networks. Although the use of directional antennas mitigates the hidden terminal problem, the protocol is not compatible with networks employing omnidirectional antennas. Sugiyama \emph{et al.} \cite{DAFD-MAC} designed a directional asynchronous FD-MAC protocol for mitigating collisions in multi-hop FD wireless networks, \textcolor{black}{however the protocol is not applicable to the omni-directional antennas, which are widely used in handheld devices}. Ramirez and Aazhang \cite{opt_pwr_rout} addressed the problem of joint power allocation and routing in FD wireless networks through a modification to Dijkstra's algorithm.  \textcolor{black}{However, the paper assumes that an FD MAC is in place. Besides, the main focus of the paper is  system-level analysis. It is also important to mention that most of the existing studies do not fully exploit the key opportunities provided by FD technology, which have been discussed later}.  On the other hand, power-aware routing protocols \cite{energyeffsurvey} for conventional HD wireless networks have received significant attention over the last few years. It can be easily inferred that  design of routing protocols for FD wireless networks requires further investigation from various aspects, which motivates this work.}

\textcolor{black}{Our objective in this paper is to design a cross-layer aided routing protocol for imperfect FD wireless networks, where the notion of imperfection implies that SI is not fully cancelled at the PHY layer. The proposed protocol, which is termed as  \textsf{X-FDR}, is particularly designed for  minimizing energy consumption and end-to-end latency in FD wireless networks. The key features of \textsf{X-FDR} can be summarized as follows. First, \textsf{X-FDR} accounts for residual self-interference (RSI) at the PHY layer. \textcolor{black}{Second, \textsf{X-FDR} adopts an optimized (not necessarily optimal) MAC protocol that implements a power control mechanism without creating the hidden terminal problem.} Third, \textsf{X-FDR} adopts a novel energy cost metric and exploits the opportunities provided by the FD technology e.g., the ability to sense the medium while transmitting. This provides immediate reaction to channel errors, and consequently, nodes are able to send a burst of packets, constrained by the minimum buffer size (\(\beta_{min}\)) on the selected route. Moreover, nodes wait for the acknowledgement (ACK) of the last received packet only instead of acknowledging from the reception of each individual packet. Fourth, nodes in \textsf{X-FDR} employ immediate forwarding, which is enabled by their FD capabilities.  A node does not have to wait for the reception of the full packet before it can forward it to the next hop. This feature reduces the end-to-end latency of the network. Last, but not the least, \textsf{X-FDR} employs a novel route maintenance process that reduces the latency due to new route discovery. Performance evaluation demonstrates that \textsf{X-FDR} provides a viable solution for multi-hop FD wireless networks.  }

\section{Opportunities of FD at Network Layer}
\label{oppo}
In this section, we describe the key opportunities provided by the FD technology that could potentially be exploited by network layer protocols. 
\begin{itemize}

\item \textbf{Immediate Forwarding} -- FD technology enables simultaneous transmission and reception, which is particularly attractive in multi-hop wireless networks. When a FD node starts receiving a packet, it can simultaneously start forwarding it to the next hop. This provides a paradigm shift from conventional \emph{store-and-forward} architecture in legacy HD networks, to \emph{receive-and-forward} architecture. For example, consider the scenario depicted in Fig. \ref{RouteExample}.  With immediate forwarding, node \(A\) can start transmitting the packet, which is being received from node \(S\), to next hop, as soon as it has processed the packet header. Immediate forwarding is particularly attractive to reduce end-to-end latency and improve throughput in multi-hop wireless networks. 

\item \textbf{Continuous Sensing} -- Another key advantage of FD technology is the ability to sense the medium while transmitting. In conventional HD networks, a node will not be notified of transmission errors, until after the transmission is complete. With continuous sensing, FD nodes can detect an erroneous transmission as soon as it occurs, which leads to immediate termination of a transmission. This improves resource utilization and potentially enables reduction of end-to-end latency. 

\item \textbf{Burst Transmission} -- The continuous sensing property further enables FD nodes to send burst of data packets, such that only the last packet is acknowledged. This is unlike conventional HD networks where packets are sent sequentially and each packet needs to be individually acknowledged. If properly exploited at the network layer, this feature has the potential to not only reduce end-to-end latency, but also improve resource utilization (particularly for signaling resources) and throughput. 

\item \textbf{Faster Convergence} -- The above mentioned features, especially immediate forwarding, enable faster dissemination of signaling information associated with routing protocols. Hence, faster topological convergence can be achieved, especially for those routing protocols that rely on building a topology tree of the network. Besides, these features can also enhance the efficiency of flooding-based routing protocols. 

\item \textbf{Secure Routing} -- Having two simultaneous transmissions on the same frequency makes it difficult for a nearby node to perform eavesdropping attacks as the received signal would be a scrambled mix of both signals. Hence, such attacks on intermediate nodes become significantly more complex to perform, thereby enhancing the security of the routing protocol between source and destination nodes.

\end{itemize}

\textcolor{black}{It is emphasized that some of the key opportunities like immediate forwarding, continuous sensing, and burst transmission have been exploited in the design of \textsf{X-FDR}. These opportunities have been further explained while discussing the protocol operation.}

\section{Network Model}\label{net_mod}
\begin{figure}
\centering
\includegraphics[scale=0.35]{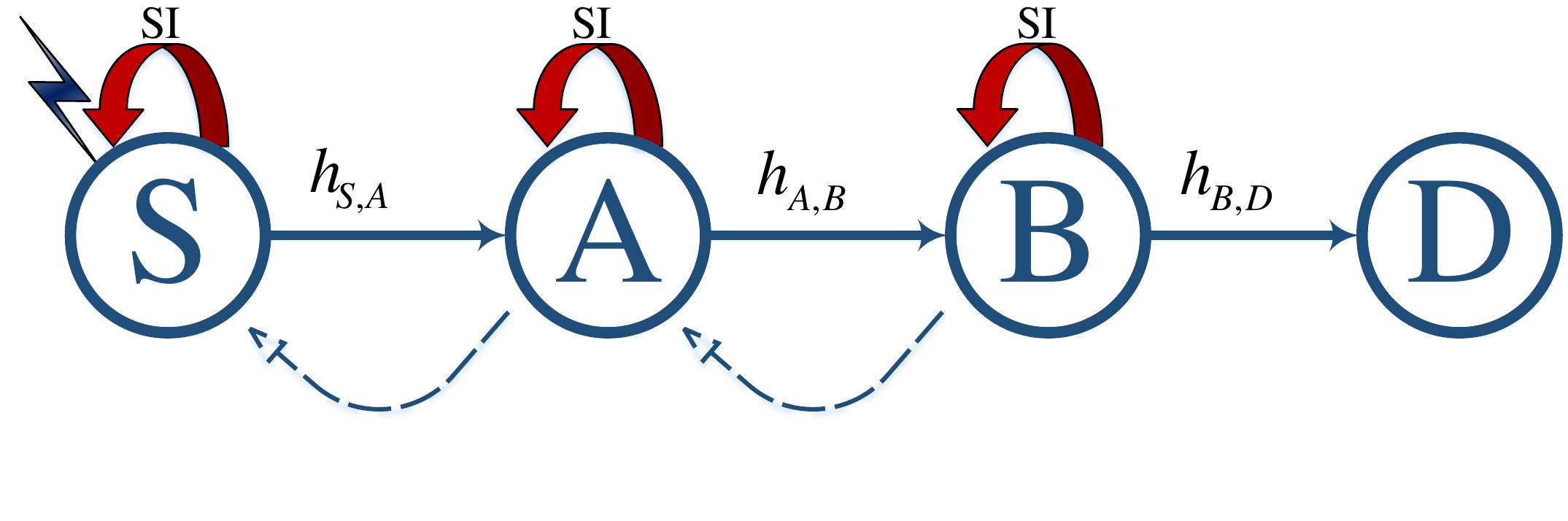}
\caption{Example of a route $R=\{S,A,B,D\}$. Straight lines represent the intended transmission, while dotted lines represent neighbouring interference, and the red semi-circled arrows represent SI.}
\label{RouteExample}
\end{figure}

\textcolor{black}{We consider a distributed network comprising \(N\) FD wireless nodes indexed by the set \(\mathcal{N}\).  Let, \(\mathcal{R}\) denote the set of all possible routes in the network. A route \(R \in \mathcal{R} \) represents an ordered set of nodes between a source node \(S\) and a destination node \(D\). For example, Fig. \ref{RouteExample} demonstrates a route comprising four nodes in the network.}

\textcolor{black}{We assume that FD wireless nodes employ necessary SI cancellation techniques at the PHY layer. Since SI cancellation techniques are not perfect in practice, a node experiences RSI.  We use an experimentally characterized model \cite{duarte2012experiment} for RSI, based on which, the power of the RSI signal is given by \( \frac{P_t^{(1-\rho)}}{\Delta \cdot \chi^\rho}\), where  \(P_t\) is the transmit power, \(\Delta\) is the interference suppression factor, \(\chi\) depends on the SI cancellation technique, and \(\rho\) denotes the SI cancellation capability. Note that \(\rho=\infty\) denotes perfect SI cancellation, resulting in zero RSI. Moreover, \(\rho=0\) implies a constant reduction in transmission power. Realistically, \(0 < \vert        \rho \vert < 1\); with \(\rho=1\) implying a constant power, for RSI similar to noise.   }

\textcolor{black}{We assume that the received signal power at a node \(j\), based on a transmission from a node \(i\) at maximum transmit power \(P_{max}\) is given by \(P_r= P_{max} \cdot \vert h_{i,j} \vert^2 \cdot d^{-\alpha}_{i,j},\) such that \(h_{i,j}\) is the channel coefficient that accounts for small-scale fading, \(d_{i,j}\) denotes the distance, and \(\alpha\) denotes the path loss exponent.} We assume that nodes in the network employ a power control mechanism based on the received signal strength such that the controlled power level is determined by \(P_{ctrl}=P_{max} \cdot (P_r)^{-1} \cdot \zeta^{th} \cdot \hat{c}\), such that \(\zeta^{th}\) denotes the minimum required received signal strength and \(\hat{c}\) is a constant \cite{Power_Control_MAC}. \textcolor{black}{Please note that RSI is not part of $P_{ctrl}$ as FD communication is not yet initialized. The impact of RSI and cumulative interference is captured on link-level}. \textcolor{black}{Further, our link-level model is based on signal-to-interference-plus-noise-ratio (SINR) which accounts for RSI and given by 
\begin{equation*}
SINR = \frac{{{P_i}\vert{h_{i,j}}\vert^2{d_{i,j}}^{ - {\alpha}}}}{{RSI+ I_x+ {N_0}}},
\end{equation*}
 where $P_i$ denotes the transmit power of node $i$ (either $P_{max}$ or $P_{ctrl}$) and $N_0$ denotes the noise power. Moreover, $I_x$ is the cumulative interference from neighboring nodes and is given by \(I_x=\sum\limits_{\textcolor{black}{x \in \mathcal{N} \setminus \{i,j\}}} {P_x} {|{h}_{x,i}|^2}{d_{x,i}}^{-\alpha}\), where $P_x$ is the transmitting power of an interfering node $x$, ${h}_{x,i}$ is the  channel coefficient between nodes $x$ and $i$, and ${d}_{x,i}$ is the distance between nodes $x$ and $i$.}

\label{MACSection}
\begin{figure}
	\centering
	\includegraphics[scale=0.5]{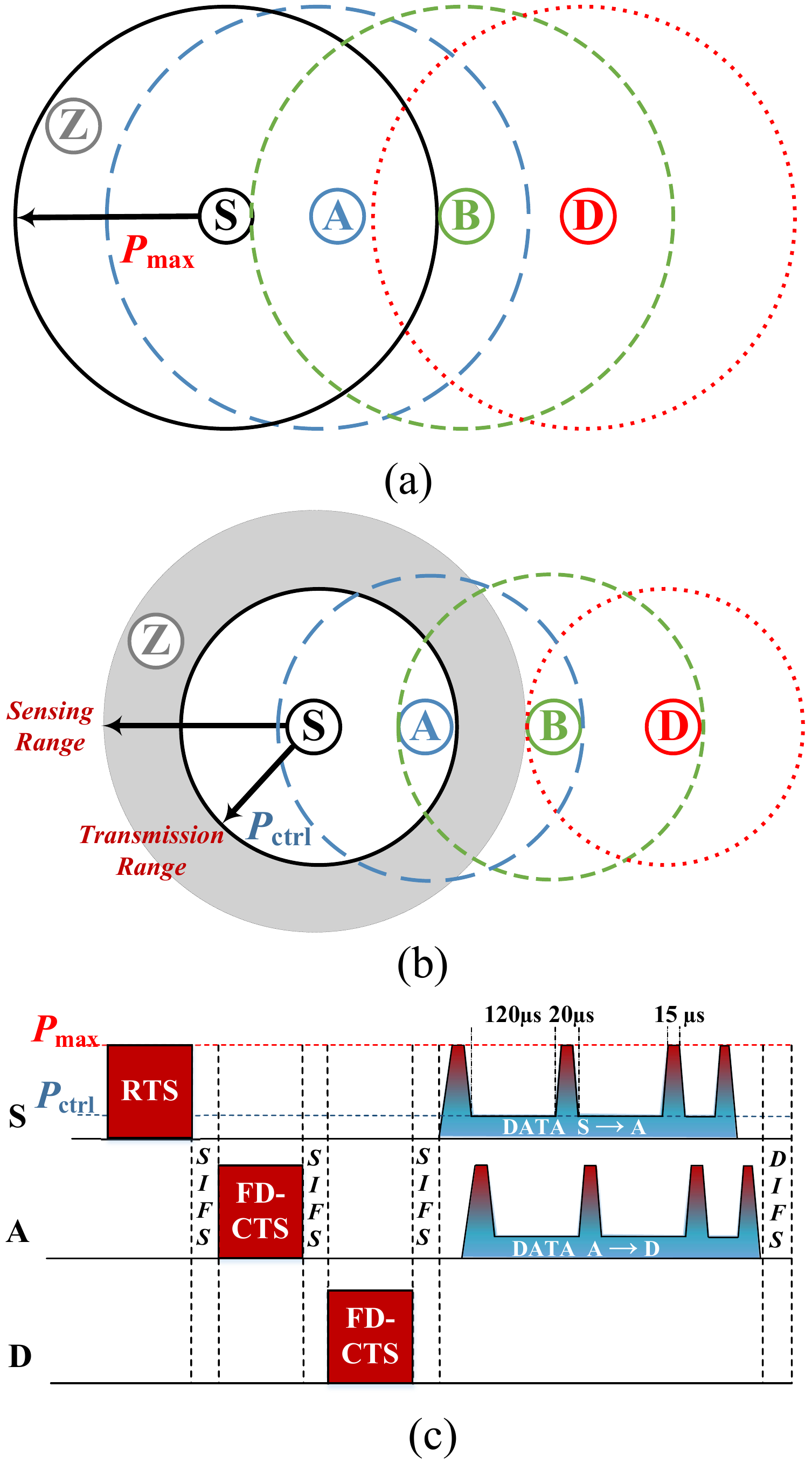}
	\caption{(a) Ranges of nodes transmitting control signals using \(P_{max}\); (b) ranges of nodes after application of power control; (c) illustration of a uni-directional FD transmission at the MAC layer.  }
	\label{Ranges}
\end{figure}

\section{MAC Layer Design for \textsf{X-FDR}}

\textcolor{black}{This section presents  the MAC layer design for \textsf{X-FDR}. In \textsf{X-FDR}, we adopt the modified version of our recently proposed MAC protocol \cite{distributedNetworkMAC} for distributed wireless networks. The MAC protocol in \cite{distributedNetworkMAC} enables both bi-directional FD transmissions and uni-directional FD transmissions. The former enables simultaneous two-way transfer of two distinct data streams between a pair of nodes, whereas, the latter involves three nodes and same data stream is forwarded from one node to another via an intermediate relay node. In \textsf{X-FDR}, we focus only on uni-directional FD transmission. We also omit the MAC layer ACK procedure.  }


\textcolor{black}{We explain the protocol operation with the aid of Fig. \ref{Ranges}.}
Let $N= \{S, A, B,D\}$ be a set of nodes involved in the intended transmission, where $S$ is the source node and $D$ is the destination node. After sensing the spectrum idle, node \(S\) starts the transmission by sending a request-to-send (RTS) packet to node \(A\) using $P_{max}$. After receiving the RTS packet from \(S\), node \(A\)  waits for short inter-frame space (SIFS) duration before sending an FD clear-to-send (FD-CTS) packet \cite{distributedNetworkMAC} to both \(S\) and \(B\). The FD-CTS packet includes the source and next hop addresses along with the transmission duration. Note that FD-CTS is also transmitted using $P_{max}$ to capture the channel for forwarding. Using the received RTS from  \(S\), node \(A\) calculates $P_{ctrl}$ as described in Section \ref{net_mod}. Node \(S\) calculates its $P_{ctrl}$ as well using the FD-CTS received from  node \(A\). Further, when node \(B\) receives the FD-CTS from node \(A\), it replies with FD-CTS as well, and calculates its $P_{ctrl}$ based on the received power from \(A\). After that, node \(A\) recalculates $P_{ctrl}$ based on the received FD-CTS from \(B\) and compares it with the previously calculated $P_{ctrl}$, where the higher $P_{ctrl}$ is chosen to maintain connection with both \(S\) and \(B\). Similarly, the rest of the relaying nodes attempt to acquire the channel until the intended destination is reached.
 
During data transmission,  nodes use $P_{ctrl}$ with periodical increase to $P_{max}$, so that nodes in the carrier sensing zone, which cannot successfully decode the transmission and set their Network Allocation Vector (NAV) to Extended InterFrame Space (EIFS) duration can sense the transmission. Note that the period between two successive power increase intervals must be less than the EIFS duration\footnote{ According to the IEEE 802.11n standard \cite{ieee80211}, \(15\) $\mu s$ is suitable for carrier sensing, and \(2\) $\mu s$ is adequate to increase/decrease the power level from/to $10\%$ to/from $90\%$. Therefore, a duration of 20 $\mu s$ is deemed adequate for transition of power level from $P_{ctrl}$ to $P_{max}$ and vice versa. Since EIFS is set to \(120\) $\mu s$, nodes will transmit at $P_{max}$ every \(120\) $\mu s$ for a duration of \(20\) $\mu s$, and the cumulative transmission duration is less than the EIFS duration.}. These periodic increments preserve the channel, and ensure that nodes in the carrier sensing zone will not attempt to initiate a transmission.


\subsection{Hidden Terminal Problem}

Referring to Fig. \ref{Ranges}b, consider that nodes \(S\) and \(A\) constitute a sender-receiver pair in HD mode. Node \(Z\), which resides in the carrier sensing range of \(S\) but not of node \(A\), may act as a hidden node. In FD transmission, hidden nodes may affect the reception of control signals at node \(S\). Therefore, in the proposed protocol we adopt RTS-CTS handshake mechanism. Moreover, by sending FD-CTS using $P_{max}$, the protocol ensures that nodes in the carrier sensing ranges are aware of an ongoing transmission. After power control is applied for data transmission, node \(Z\) can again create a hidden node problem, which is why the periodic increments from $P_{ctrl}$ to $P_{max}$ are required. 

\section{\textsf{X-FDR}: Protocol Operation}
\label{the_protocol}
\textcolor{black}{This section explains the protocol operation of \textsf{X-FDR}. Unlike conventional Adhoc On-demand Distance Vector (AODV) routing protocol \cite{AODV},  where the route cost relies mainly on hop count, \textsf{X-FDR} uses  energy consumption as the key metric for route cost estimation. }


\subsection{Route Cost Estimation}
 Since \textsf{X-FDR} is a cross-layer routing protocol, all relevant factors must be accounted for in route cost estimation. Nodes in the network initiate connections using RTS/FD-CTS messages with maximum power level $P_{max}$, in order to restrain other nodes residing in the sensing range from initiating an interfering transmission. Once the data transmission take place using controlled power level $P_{ctrl}$, a periodic increase of power  to $P_{max}$ takes place to stop potential interference and eliminate the problem of hidden nodes; therefore, the metric for route cost shall account for different power levels.

In a route $N= \{S, A, B,D\}$, the cost of energy for sending data from node \(S\) to node \(A\) can be estimated as \(\rho_{(S,A)}=\chi(E_{data}+E_{ctrl}+E_{on})\),
where $\chi=1/\mathcal{P}_f$ is the the number of retransmissions attempts such that ${\mathcal{P}_f}$ denotes the probability of transmission failure, and  $E_{data}$, $E_{ctrl}$ and $E_{on}$ denote the energy consumed during data transmission, control signal transmission and when the receiver is turned on, respectively. The energy consumption during data transmission phase can be calculated as \(E_{data}=P_{ctrl}^S(\beta_{min}/{r}-T_{inc})+P_{max}T_{inc}\),
where $P_{ctrl}^S$ and $P_{max}$ denote the controlled power level and maximum transmit power of node \(S\), respectively, $\beta_{min}$ is the minimum buffer size (explained in Section \ref{RD}), and $r$ is the data rate. Moreover, $T_{inc}$ denotes the duration of the periodic increase/decrease in power levels. The energy consumption during control signal transmission can be calculated as $E_{ctrl}=P_{max}(T_{RTS}+T_{FD-CTS})$,
where $T_{RTS}$ and $T_{FD-CTS}$ denote the duration of  RTS and FD-CTS messages, respectively. 

Assume that there exists a route  ${R}_i = n_o \rightarrow n_1 \rightarrow ... \rightarrow n_k$ from the source node \(S\) to the destination \(D\), where, without loss of generality, $S=n_0$ and $D=n_k$. Therefore, the total cost, $\bar{\rho_i}$,  along the route $\mathcal{R}_i$ can be expressed as

\begin{equation*}
\bar{\rho_i}=\sum_{j=0}^{k-1}\rho_{(j,j+1)}(P_j),
\end{equation*}
where $P_j$ is the power level used by node $n_j$ to communicate with node $n_{j+l}$, and $\rho_{(j,j+1)}(P_j)$ is the relaying cost between nodes $n_j$ and $n_{j+l}$. Assuming that there are $x$ routes from source to destination, the objective of the routing protocol is to select the route with minimum energy consumption i.e., $R_{min}=\argmin(\bar{\rho_i}), \quad \forall \ i=1,2,\dots,x.$ 


\subsection{Route Discovery}
\label{RD}

\begin{figure}
	\centering
	\includegraphics[scale=0.45]{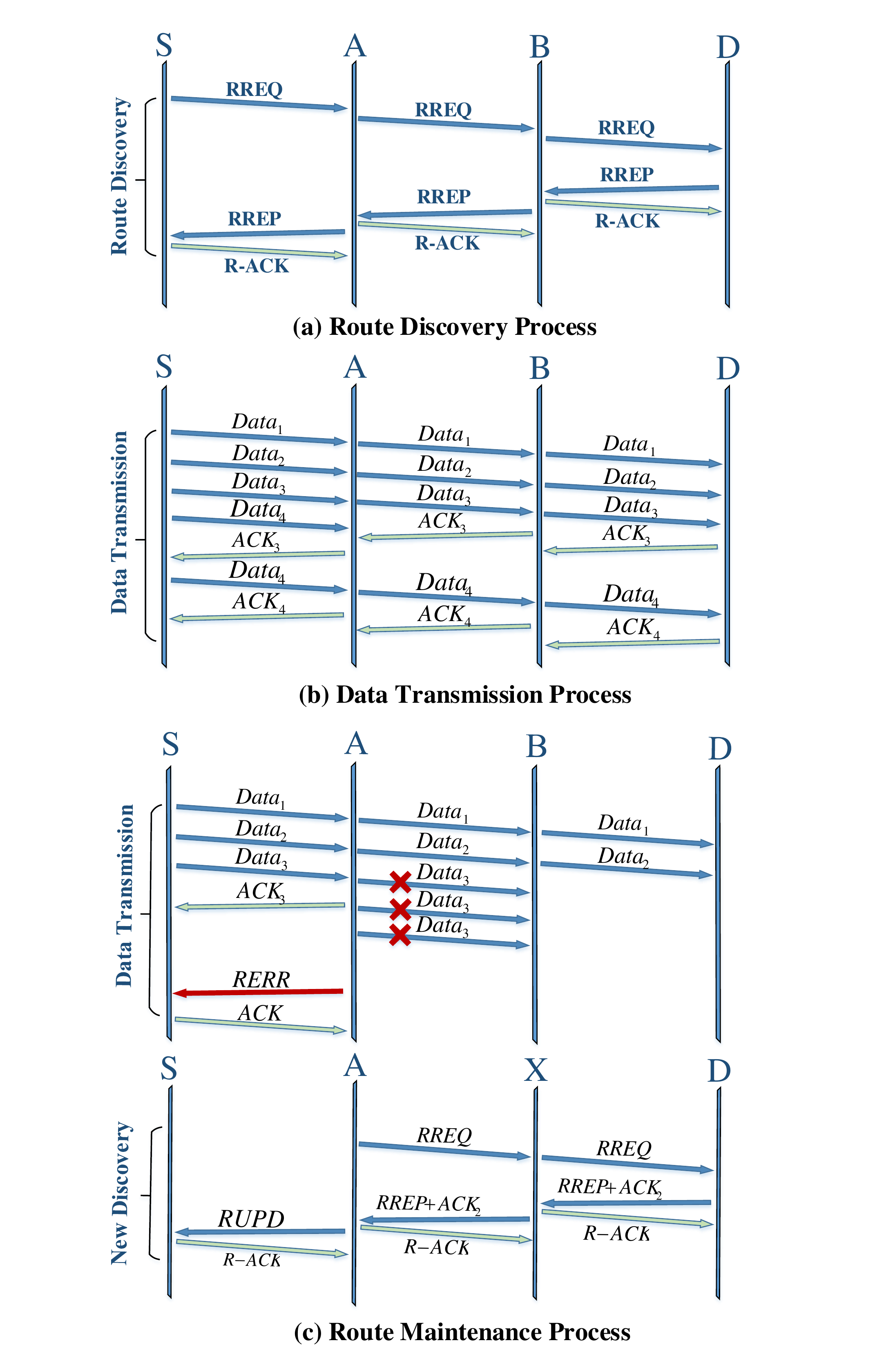}
	\caption{Example of (a): route discovery process; (b) data transmission process; (c) route maintenance process in \textsf{X-FDR}. }
	\label{Route_discovery}
\end{figure}

\textcolor{black}{The first stage of \textsf{X-FDR} is route discovery. When a source node \(S\) requires a route to destination node \(D\), it broadcasts a Route REQuest message (RREQ). Once neighbouring nodes receive RREQ, they calculate the energy cost, $\rho$, add it to RREQ and  broadcast it to the neighboring nodes. After that the neighboring nodes  calculate the new $\rho$, add it to the previous cost received in RREQ, and broadcast it further until it reaches the destination \(D\). The destination node  sets up a timer to allow several RREQ messages to arrive from different routes. After the timer expires, node \(D\) chooses the route $R_{min}$ with minimum energy consumption and replies with a Route REPly (RREP) message via $R_{min}$. The routing table of each node is refreshed, whenever it receives RREQ/RREP messages. Each node maintains a received RREQ table and compares the new RREQ messages  in order to eliminate the duplicate RREQ messages. Additionally, a Route-ACKnowledgement (R-ACK) packet is used by the nodes receiving RREP, in order to confirm successful reception of the RREP packet and establishment of the route. Fig. \ref{Route_discovery} demonstrates an example of route discovery performed by node \(S\), where $R_{min}$ is found to be $\{S,A,B,D\}$.}

\textcolor{black}{Instead of sending packets sequentially and waiting for acknowledgements (ACKs) for each data packet, \textsf{X-FDR} sends a burst of packets, such that the number of packets in each burst is determined by the minimum buffer size, $\beta_{min}$, of the nodes in the route $R$. For example, let $\beta_S$ denote the buffer size (in terms of the number of packets) of node \(S\). Further, node \(S\) encapsulates $\beta_{min}=\beta_S$ within RREQ and broadcasts it. When a neighboring node \(A\) receives RREQ, it compares $\beta_{min}$ with its own buffer size (i.e., $\beta_A$). If $\beta_A <\beta_{min}$, node \(A\) updates $\beta_{min}=\beta_A$ in RREQ and broadcasts it forward. However, if $\beta_A>\beta_{min}$, node \(A\) keeps the buffer size as it is and forwards RREQ. When \(D\) receives RREQ, it  compares its buffer size with the received $\beta_{min}$, and sends the lowest of the two within RREP, which informs node \(S\) with the minimum buffer size to be used for data transmission. Note that if a node in the route does not have a buffer enabled, $\beta_{min}$ will be set to 1.}

\subsection{Data Transmission}
When node \(S\) receives the RREP message as a result of route discovery process, it becomes aware of the most energy-efficient route to the destination \(D\). After the route discovery process, node \(S\) starts transmitting a burst of data packets to next hop (node \(A\)), where the number of packet in each burst is given by  $\beta_{min}$ of the route. 
\textcolor{black}{ In conventional HD communications, when node \(S\) sends a burst of data, it will not be notified of transmission errors, e.g., by receiving a Route ERRor (RERR) message, until after the entire burst is transmitted. This incurs significant waste of  time and resources. However, using continuous sensing offered by FD technology, node \(S\) can sense a problem in the transmission as soon as it occurs, which leads to immediate termination of the transmission. } Hence, node \(S\) continuously senses the packets forwarded by node \(A\) and stops transmitting immediately if it receives RERR message.

 Since node \(A\) is FD-capable, it can employ \emph{immediate forwarding}, wherein it does not have to wait for the entire packet to be received before forwarding. Once node \(A\) receives all the packets, determined by $\beta_{min}$, it replies with an ACK to acknowledge the reception of the last packet. If a packet is dropped while the route is not deemed faulty, node \(S\) gets notified by the ACK packet sent by \(A\), and it retransmits the lost packet. The same process is repeated at each hop until the destination is reached. Note that each node in the route only notifies the previous hop with an ACK. This is because data is assumed to be buffered by the previous node in the route as $\beta_{min}$ is known to all nodes. Therefore, if a node did not receive all  the packets, it would request these from previous nodes using ACK.

For instance, assume that $\beta_{min}=4$, and consider the scenario  demonstrated in Fig. \ref{Route_discovery}. Node \(S\) transmits data packets 1 through 4 while continuously sensing the signal transmitted by \(A\). Node \(A\) starts forwarding immediately; however it only receives 3 packets.  Therefore, it sends an ACK for data packet 3, which notifies \(S\) that it needs to retransmit data packet 4. 
\textcolor{black}{Note that if the buffer size of node \(S\) is larger than the amount of data packets that needs to be sent, it will include an end-of-queue (EQ) notification message with the last packet, in order to avoid an unnecessary retransmission.   }

\subsection{Route Maintenance}
\textcolor{black}{The process of route maintenance is depicted in Fig. \ref{Route_discovery}, where the transmission of packets 1 to 3 from source \(S\) to  destination \(D\) is exemplified. First, node \(S\) transmits the burst of packets to node \(A\). Node \(A\) receives the packets successfully  and responds with an ACK packet to the source \(S\) to confirm the successful reception. As node \(A\) receives the packets, it starts forwarding them to node \(B\). However, node \(B\) fails to receive the data packet 3 successfully, despite maximum number of retransmission attempts by node \(A\) due to a link error. Once the pre-set timer expires at node \(A\) without receiving any ACK from node \(B\), it infers that  the link \(A\--B\) is broken and sends an RERR message to its previous hop, which is node \(S\) in this case. The RERR message informs node \(S\) of a link failure and a new route discovery process. Node \(S\) updates its routing table and marks link \(A\--B\) as broken, and then acknowledges the RERR of node \(A\). Since the route error occurred at node \(A\), it initiates a new route discovery process by broadcasting a RREQ message. Intermediate nodes follow the same procedure as described earlier for route discovery. When the destination node \(D\) receives RREQ from node \(A\), prior to the full reception of packets in the same burst of $\beta_{min}$, it knows that the request is to complete the same data stream, and replies with RREP, piggybacked with an ACK packet to inform node \(A\) about the last packet node \(D\) had received. After receiving the RREP message, node \(A\) sends a Route UPDate (RUPD) message, with the new $\beta_{min}$, to the previous hop i.e., node \(S\), to inform it of a new route. Finally, node \(A\) starts new data transmission to destination \(D\) from data packet 3 onwards. If node \(S\) has new burst of data to send, it will use the updated route towards node \(D\), starting the transmission after sensing the last packet sent by node \(A\). The process of route maintenance is summarized in Algorithm \ref{routemain}.} \qed

\textcolor{black}{\emph{Remark 1} -- It is worth emphasizing that \textsf{X-FDR} incurs less overhead and complexity as compared to its HD counterparts. First, it omits the MAC layer ACK procedure which reduces the signaling overhead. Second,   from the routing perspective, the overhead in most cases is reduced which simplifies the system design. For instance, in the route discovery process, ACK packets are only sent to acknowledge the RREP packets which reduces the overhead significantly as compared to acknowledging the RERR packets. Similarly, acknowledging a burst of packets instead of each packet reduces the overhead in the network. }

\textcolor{black}{\emph{Remark 2} -- Some recent studies \cite{air_exp, bpass}  have investigated the problem of \emph{in-band wireless cut-through} which is closely related to the problem of multi-hop transmissions in FD wireless networks. To realize wireless cut-through transmissions, specialized hardware is required for cancellation of all types of interference. It is worth emphasizing that the need for MAC and routing protocols cannot be eliminated for realizing wireless cut-through transmissions. \textsf{X-FDR} adopts a cross-layer approach for multi-hop transmissions in FD wireless networks and focuses only on SI cancellation which can be achieved through state-of-the-art FD radios. \textsf{X-FDR} can directly benefit from additional hardware capabilities as realized for wireless cut-through transmissions. Alternatively, the cross-layer approach of \textsf{X-FDR} can improve the efficiency of wireless cut-through solutions. 
}


\begin{algorithm}
\DontPrintSemicolon
\KwIn{Source: $S$, Destination: $D$, Nodes: $N$, $R_{min}$}
\KwOut{New Route: Updated $R_{min}$}
\While{$S\rightarrow D$ }{Nodes forward incoming packets \\
\For {each node $i$ $\in \{R_{min} \setminus D\}$}
{
$i$ transmits packets to $i+1$\\
$i$ sets timer $t_{ACK}$\\
\lIf{ $i$ receives ACK${}_z$ while $t_{ACK}$$\neq$0 }{\\ $i$ forwards packets $z+1$}
\lElse{ $i$ marks link $i \rightarrow (i+1)$ as broken\\
Send RERR to node $i-1$\\
Nodes $i-(x+1)$, $x \in \{0, 1, \dotsi, \text{hops to \(S\)}\}$ traverse RERR back to \(S\)\\
$i$ broadcasts RREQ\\
\If{ D receives RREQ for the same stream}{set a timer $t_{max}$ \\ \lIf{$t$ $\le t_{max} $ }{\\ continue receiving RREQ packets}\lElse{ stop receiving RREQ packets; \\compare received RREQ packets and select  $R_{min}$\\\Return   RREP packet with $R_{min}$ and $\beta_{min}$.\\
\If{ $i$ receives new RREP}{Update $R_{min}$\\Send RUPD to node $i-1$\\Nodes $i-(x+1)$, $x \in \{0, 1, \dotsi, \text{hops to \(S\)}\}$ update $R_{min}$ and send RUPD back to \(S\)\\ \textcolor{black}{Closest node to \(D\) with full $\beta_{min}$ received will resume transmission}}}}}}
}
\caption{Route Maintenance Process in \textsf{X-FDR}}
\label{routemain}
\end{algorithm}

\section{Performance Evaluation}
\label{simulation}
\textcolor{black}{In this section, we conduct a performance evaluation of \textsf{X-FDR}. We have implemented \textsf{X-FDR} in OPNET. Necessary changes were made in the node and protocol models to implement simultaneous transmission and reception. We assume that nodes are randomly distributed in an area of \(500 \ \text{m}^2\). The buffer size is assumed to be fixed and set to \(10\) kB. The maximum transmit power of a node is set to \(23\) dBm (\(200\) mW). We assume a channel bandwidth of \(2\) MHz. The path loss exponent is set to \(4\). We consider file transfer protocol (FTP) application with packet size of \(1\) kB. The RSI parameter, \(\chi\) is set to \(13\) dB.  The simulation results are averaged over \(10\) iterations. In each iteration, source and destination nodes are randomly selected. \textcolor{black}{We have modified the wireless model in OPNET to account for RSI and Rayleigh fading. }
For performance comparison, we select two different baseline protocols: AODV and FD version of AODV, termed as FD-AODV, wherein nodes employ immediate forwarding and acknowledge each packet. Moreover, both AODV and FD-AODV do not employ power control.}

\begin{figure*}
\centering
\subfloat[]{\includegraphics[scale=0.29]{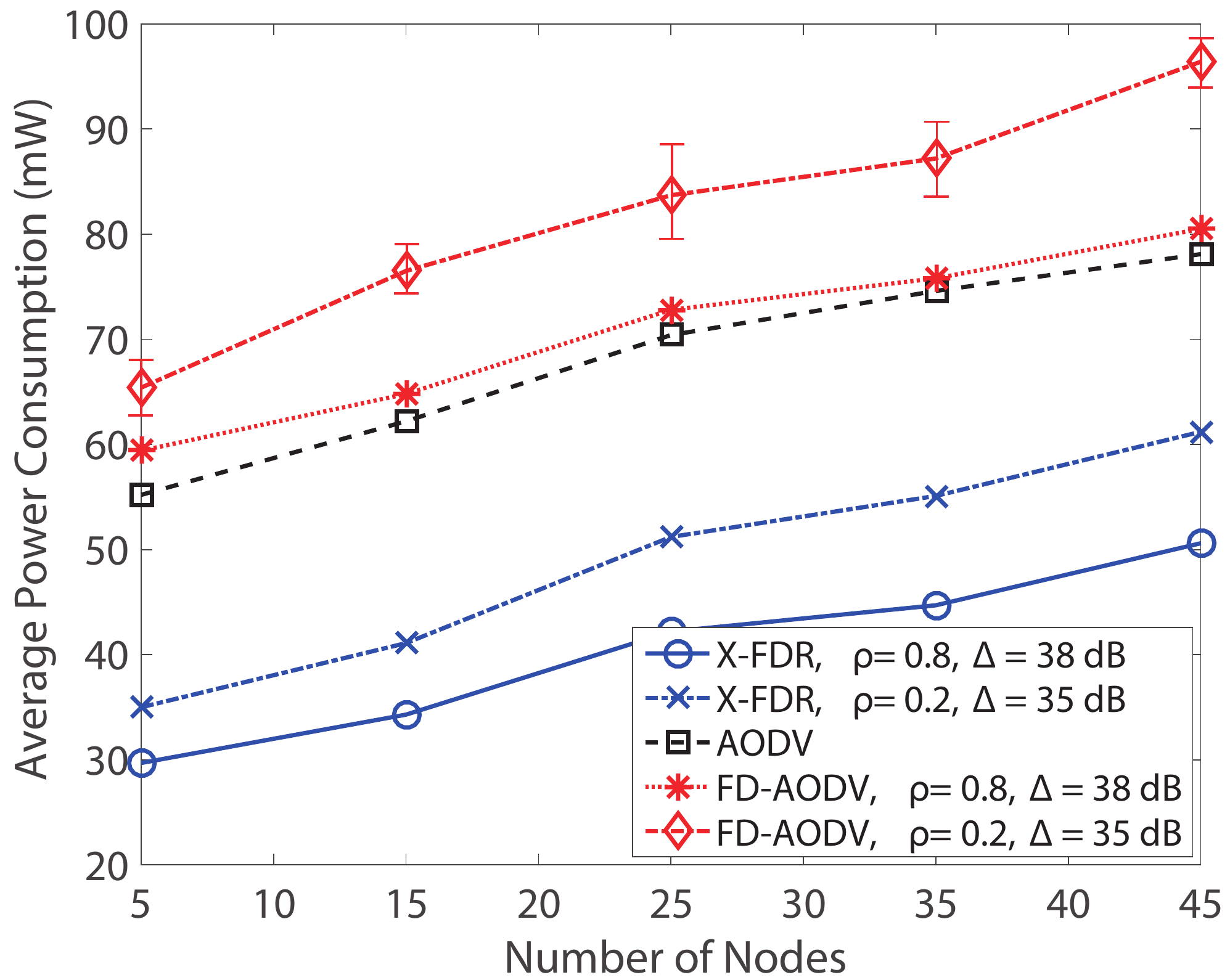}\label{NodesVsPower}}\
\subfloat[]{\includegraphics[scale=0.29]{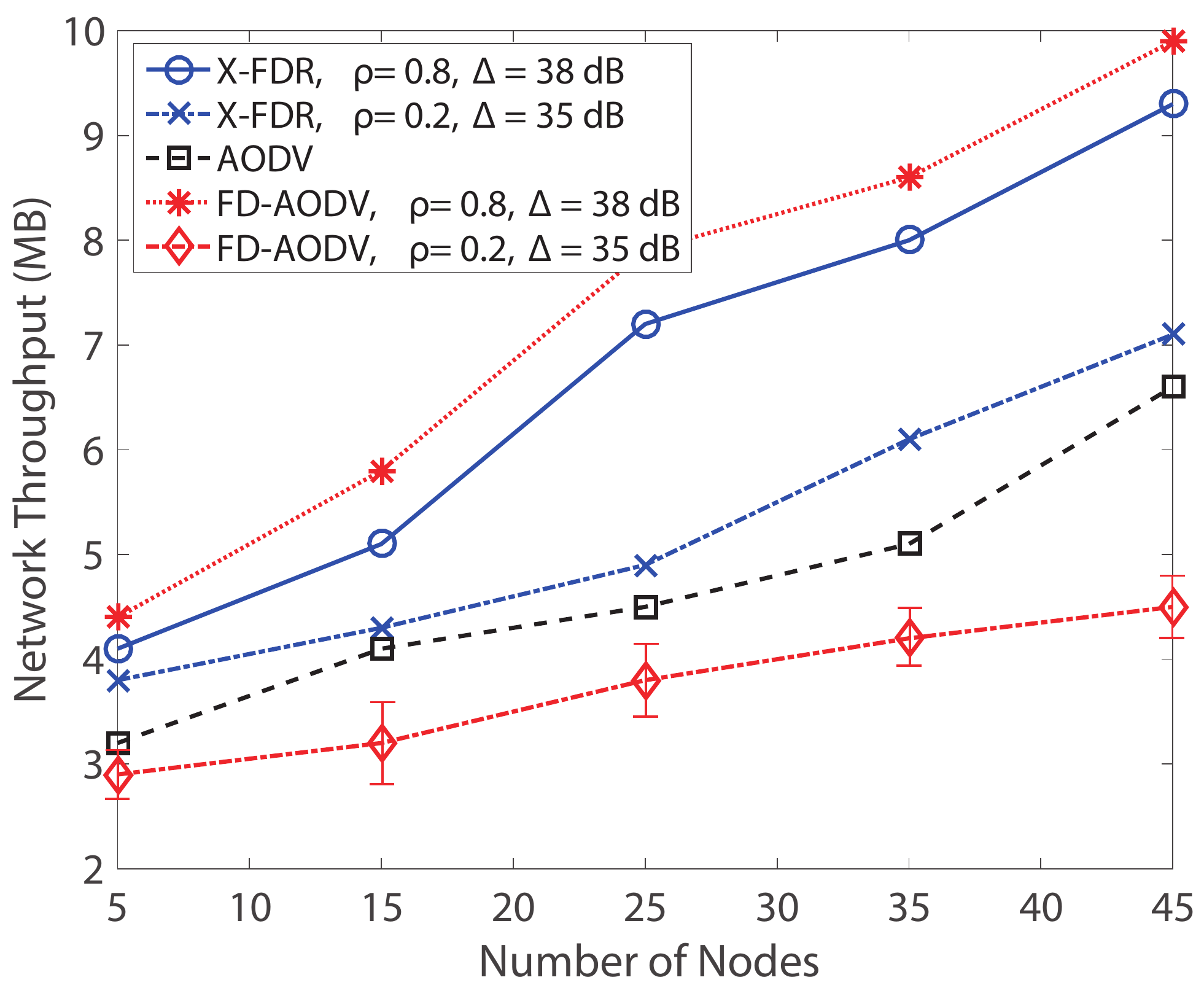}\label{NodesVsThroughput}}\
\subfloat[]{\includegraphics[scale=0.29]{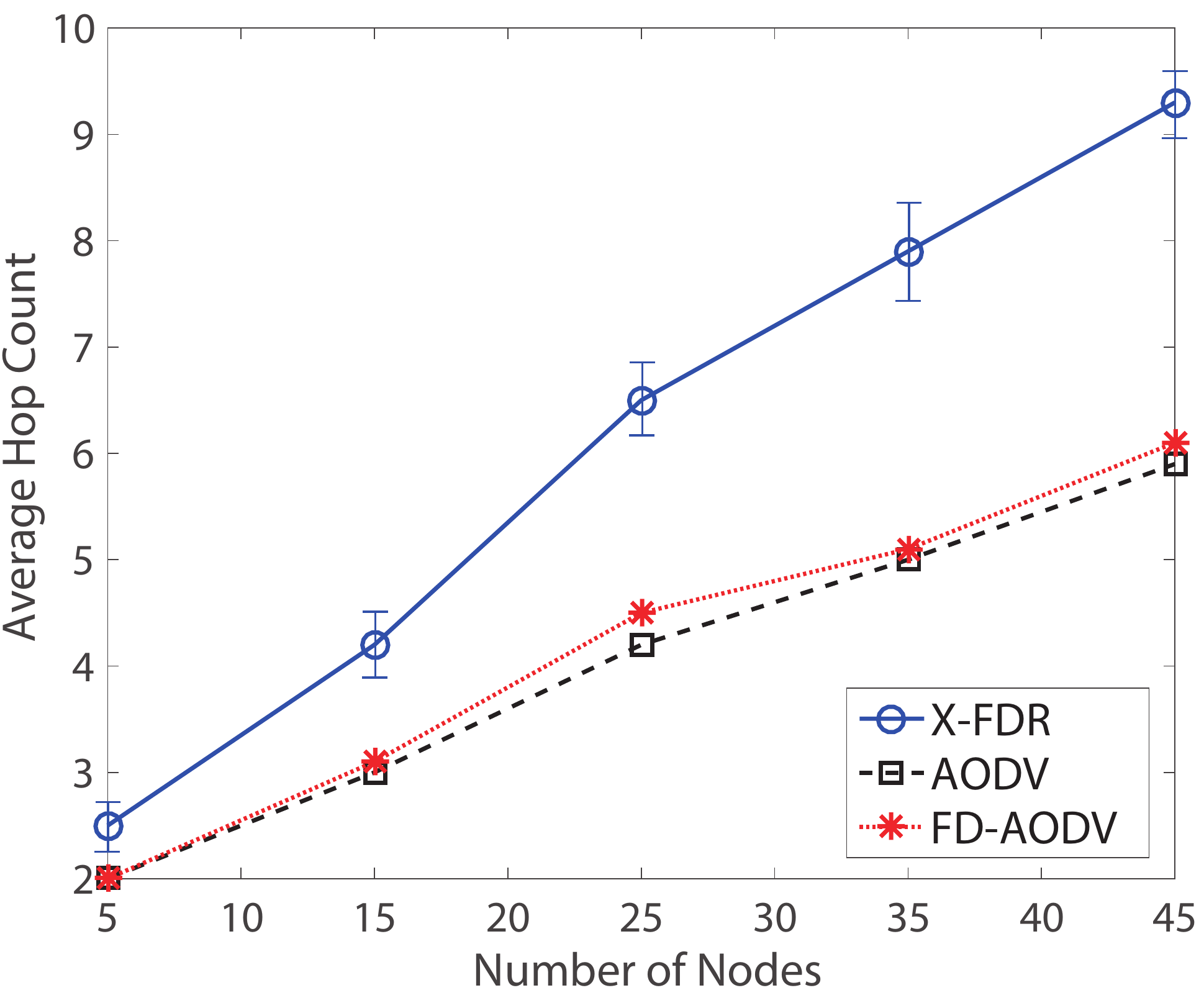}\label{NodesVsHopCount}}
\\
\subfloat[]{\includegraphics[scale=0.285]{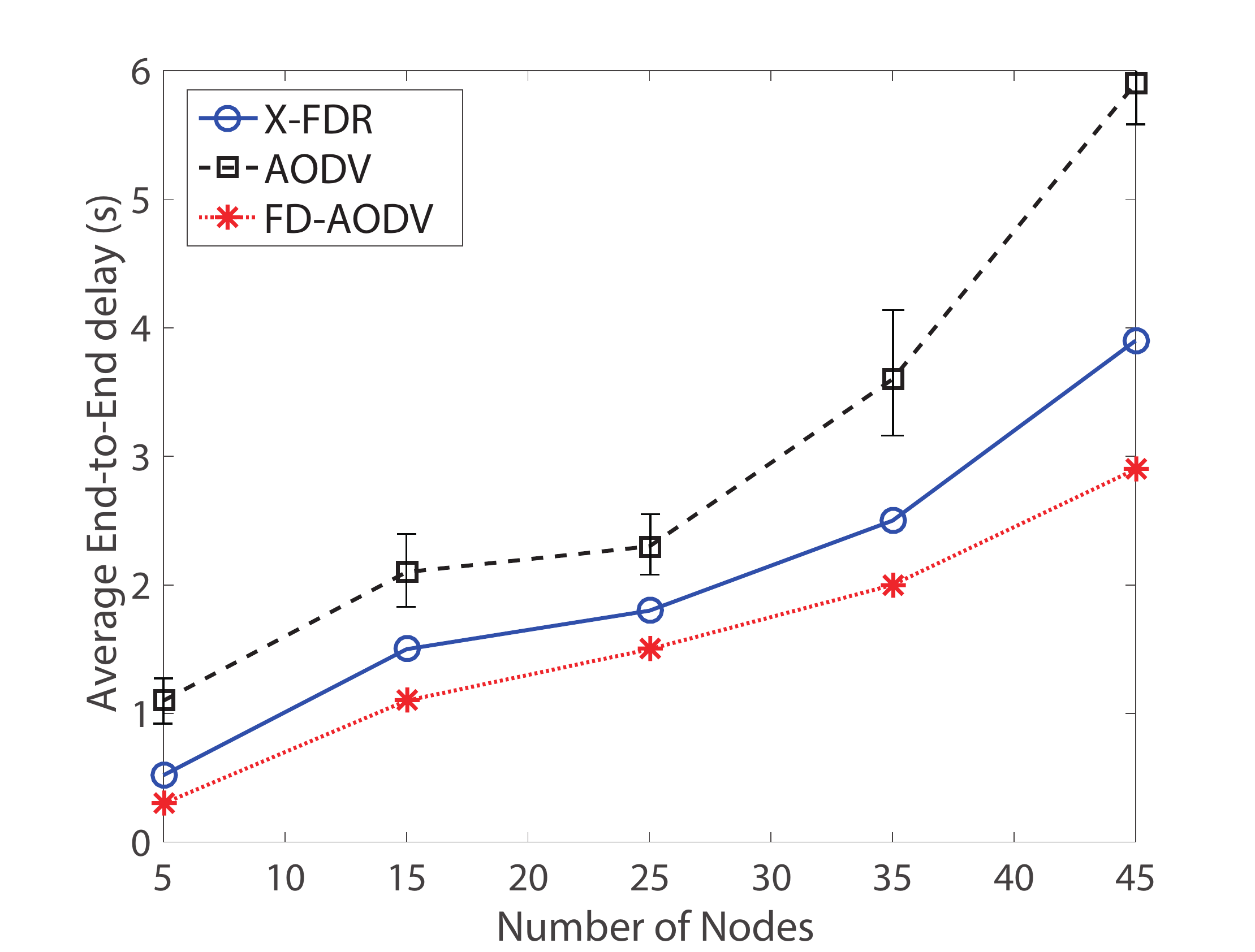}\label{NodesVsAEndtoEndDelay}}\
\subfloat[]{\includegraphics[scale=0.285]{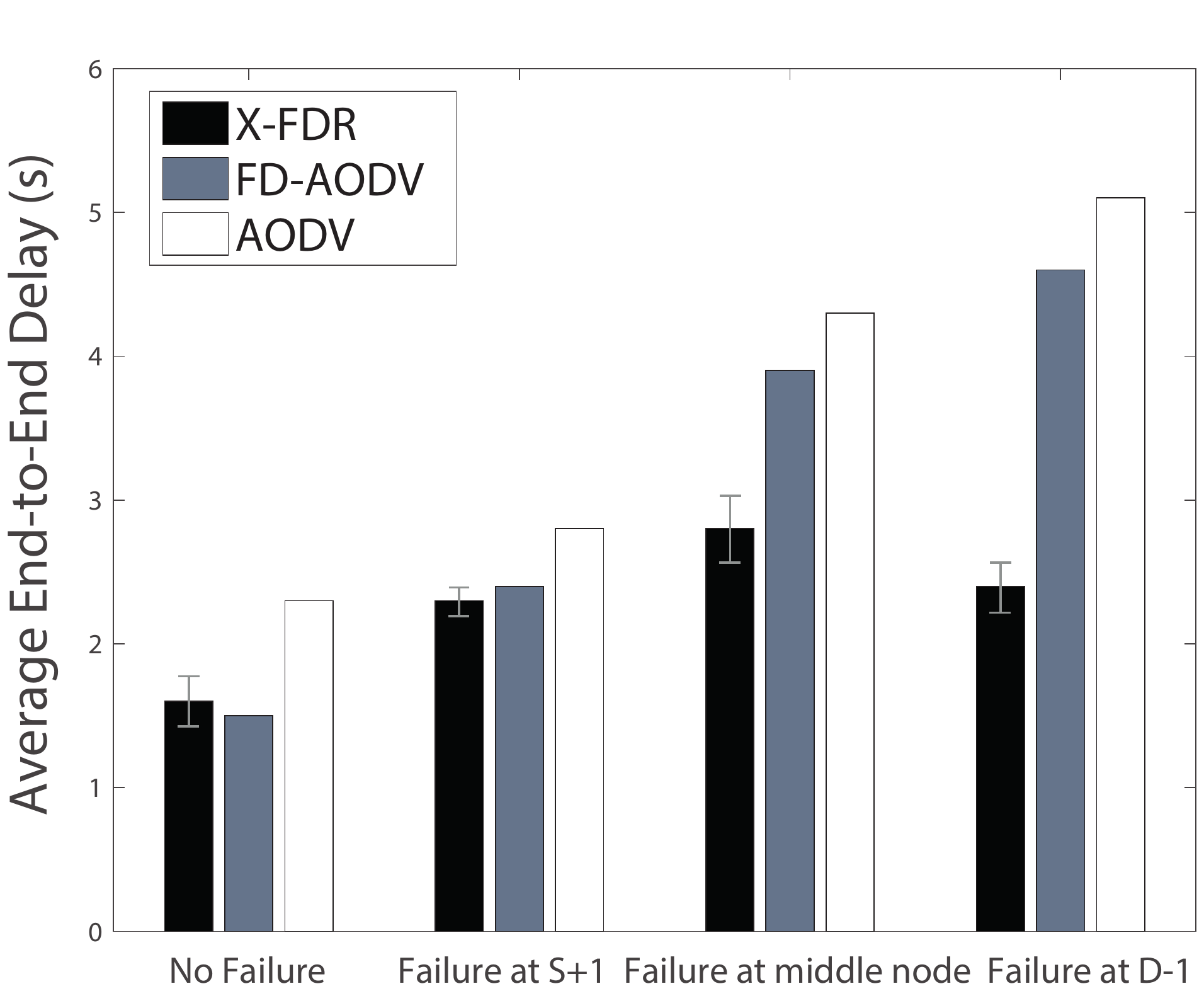}\label{NodesVsEnd-to-EndDelay}}\
\subfloat[]{\includegraphics[scale=0.285]{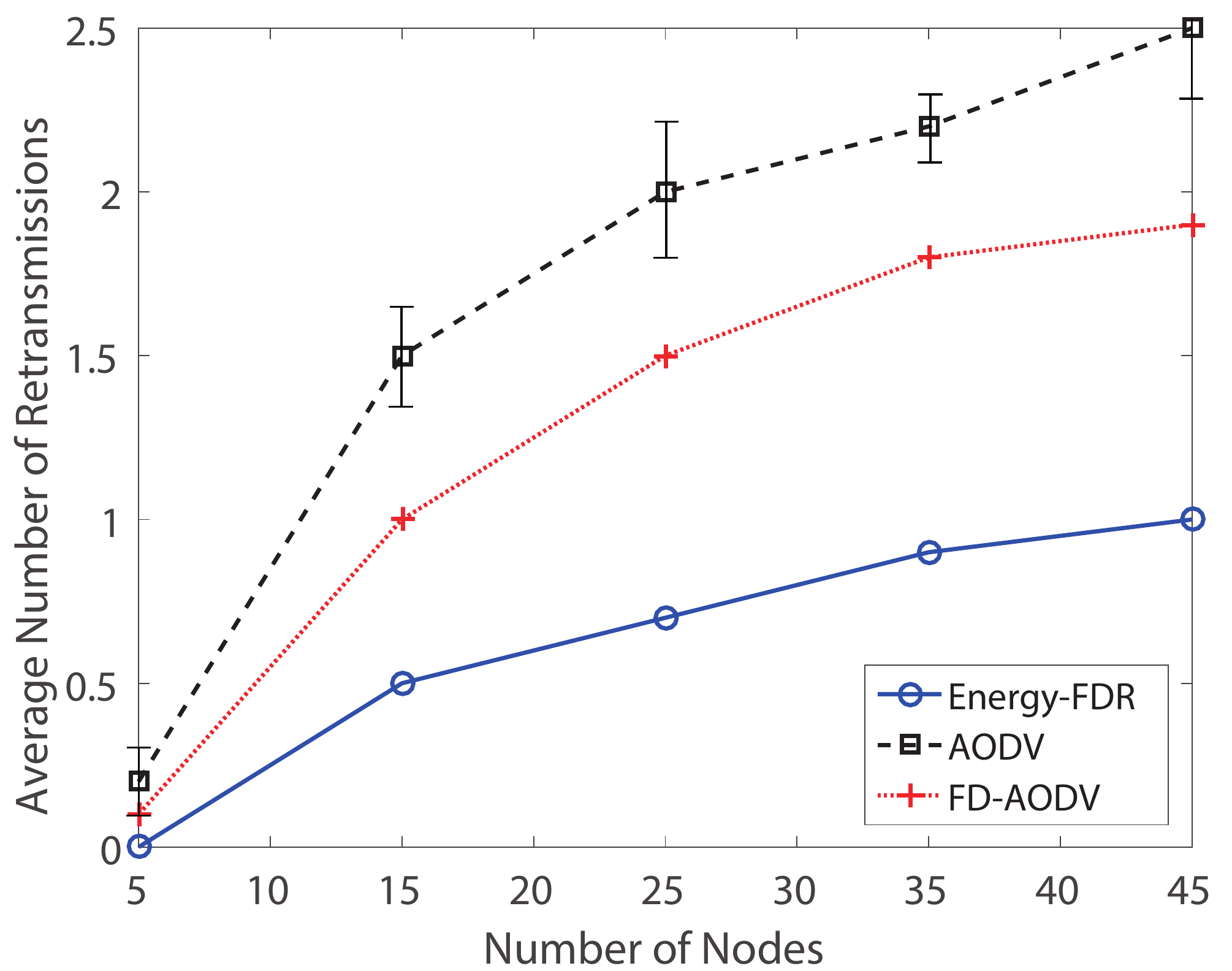}\label{NodesVsRetransmission}}
\caption{Performance evaluation of \textsf{X-FDR}: (a) average power consumption; (b) network throughput; (c) average hop count; (d) average end-to-end delay; (e) average end-to-end delay with node failures (number of network nodes = \(25\)); (f) average MAC layer retransmissions. \textcolor{black}{The confidence intervals on different figures are also shown.}}
\end{figure*}

\textcolor{black}{Fig. \ref{NodesVsPower} shows the average power consumption of routes from source to destination nodes selected by different protocols. First, we note that the power consumption increases with the number of nodes in the network. This is due to inclusion of more nodes in the routes selected by different protocols. Second, we note that \textsf{X-FDR} outperforms both baseline protocols by performing up to 40\% and 50\% better than AODV and FD-AODV protocols, respectively. The performance gain of \textsf{X-FDR} in terms of energy-efficiency is primarily due to the use of power control at the MAC layer, which limits the effect of interference, and the adoption of energy-based routing cost metric. Third, we note that SI cancellation plays an important role in power consumption. A higher SI cancellation capability, corresponding to  higher values of \(\Delta\) and \(\rho\), reduces the power consumption due to less number of transmission failures due to interference.}

\textcolor{black}{Fig. \ref{NodesVsThroughput} shows network throughput against the number of network nodes. We note that \textsf{X-FDR} outperforms AODV by performing up to 50.2\% and 21.2\% better under high and low SI cancellation scenarios, respectively. This is primarily due to the FD features of \textsf{X-FDR}. Further, \textsf{X-FDR} achieves nearly 8.6\% lower throughput than FD-AODV under high SI cancellation scenario. This can be attributed to the employment of power control in \textsf{X-FDR} as there is an inherent trade-off between power and throughput. Note that the presence of SI, due to low SI cancellation capability, can degrade the performance of FD-AODV to the extent that it achieves lower throughput than AODV. Such performance degradation is also visible in case of \textsf{X-FDR}.  }


\textcolor{black}{Fig. \ref{NodesVsHopCount} plots the average hop count between randomly located source and destination nodes, as a function of number of nodes in the network. The average hop count increases with the number of network nodes as more nodes are involved in the selected routes.\textcolor{black}{ We note that \textsf{X-FDR} has higher average hop count than the baseline protocols. This is because both AODV and FD-AODV use hop count as the routing metric. However, \textsf{X-FDR} focuses on routes with minimal energy consumption, and therefore, it incurs higher hop count with lower total energy consumption.}}

\textcolor{black}{Fig. \ref{NodesVsAEndtoEndDelay} plots the average end-to-end delay against the number of network nodes. We note that \textsf{X-FDR} outperforms AODV by achieving up to \(33\%\) lower delay, due to the use of immediate forwarding, continuous sensing and burst transmission mode. On the other hand, FD-AODV outperforms \textsf{X-FDR} by achieving up to 12\% lower delay. This is due to the fact that \textsf{X-FDR} incurs higher hop count. Although both AODV and FD-AODV incur similar hop count, the latter achieves lower delay due to immediate forwarding feature.  It is important to mention here that the results in Fig. \ref{NodesVsAEndtoEndDelay} correspond to the scenario  when the route does not suffer any failures along its path. In order to capture the impact of route maintenance, we deliberately mark nodes to fail (an option provided by OPNET) across the route during transmission process and evaluate end-to-end delay in Fig. \ref{NodesVsEnd-to-EndDelay}. 
Initially, we fail the first node after the source, then a node at the middle of the route, and finally, a node right before the destination for worst case scenario. As shown by the results, \textsf{X-FDR} outperforms both AODV and FD-AODV by performing up to 39\% and 34\% better than the former and the latter, respectively. The performance gain is due to the proposed route maintenance procedure that initiates a route discovery process at the last buffered node instead of starting new route discovery process by the source. }

 Fig. \ref{NodesVsRetransmission} shows the average number of MAC layer retransmission attempts against the number of network nodes. The average number of retransmissions increase with the number of network nodes due to higher probability of failures as a result of higher inter-node interference. We note that \textsf{X-FDR}  incurs the lowest number of retransmissions than both AODV and FD-AODV. This is primarily due to an optimized MAC protocol that minimizes collisions due to hidden node problem while using power control. 

Finally, a qualitative comparison of \textsf{X-FDR} against state-of-the-art protocols is given in \tablename~\ref{qual_comp}.

\begin{table*}
		\caption{Qualitative Comparison of Different Routing Protocols for FD Wireless Networks}
		\begin{center}
			\begin{tabular}{lcccccc}
				\hline	
				\toprule

				\textbf{Feature\(/\)Protocol} &  \textbf{OMR}\cite{distributedFD}   &  \textbf{D-FDW}\cite{directionalFD} & \textbf{M-DA} \cite{opt_pwr_rout} & \textbf{AODV} &  \textbf{FD-AODV} & \textbf{\textsf{X-FDR}}   \\ \hline
			\midrule			
				Residual SI & No & No & Yes & No & Yes & Yes  \\
				Power Control & No & No & No & No & No & Yes    \\
				Directional Antennas & No & Yes & No & No & No & No   \\
				Optimized MAC & Yes & Yes & No & No & No & Yes   \\
				Immediate Forwarding & No & Yes & No & No & Yes & Yes   \\
				Continuous Sensing & No & No & No & No & No & Yes   \\
				Burst Transmission & No & No & No & No & No & Yes   \\
				Energy Efficiency & No & No & No & No & No & Yes  \\
				\hline
								
			\end{tabular}
		\end{center}
			\label{qual_comp}
	\end{table*}

\section{Concluding Remarks}
\label{conclusion}
\textcolor{black}{FD technology has the potential to play an important role in realizing the capacity objectives of future wireless networks.  Realizing the FD capability at higher layers of the protocol stack is particularly attractive to reap the full potential of FD technology. In this paper, we have designed a cross-layer routing protocol, termed as \textsf{X-FDR}, for multi-hop FD wireless networks with imperfect SI cancellation. \textsf{X-FDR} accounts for RSI at the PHY layer, adopts an optimized MAC protocol with power control feature, and exploits the opportunities provided by FD technology at the network layer. Performance evaluation demonstrates that \textsf{X-FDR} outperforms baseline protocols in terms of power consumption without a significant compromise on network throughput. Besides, it achieves lower end-to-end delay in the presence of route failures. Hence, \textsf{X-FDR} provides a viable solution for multi-hop FD wireless networks.   }

\ifCLASSOPTIONcaptionsoff
  \newpage
\fi



\bibliographystyle{IEEEtran}
\bibliography{My_citation}


\bibliography{IEEEabrv,mybibfile}

\begin{thebibliography}{10}
\providecommand{\url}[1]{#1}
\csname url@samestyle\endcsname
\providecommand{\newblock}{\relax}
\providecommand{\bibinfo}[2]{#2}
\providecommand{\BIBentrySTDinterwordspacing}{\spaceskip=0pt\relax}
\providecommand{\BIBentryALTinterwordstretchfactor}{4}
\providecommand{\BIBentryALTinterwordspacing}{\spaceskip=\fontdimen2\font plus
\BIBentryALTinterwordstretchfactor\fontdimen3\font minus
  \fontdimen4\font\relax}
\providecommand{\BIBforeignlanguage}[2]{{%
\expandafter\ifx\csname l@#1\endcsname\relax
\typeout{** WARNING: IEEEtran.bst: No hyphenation pattern has been}%
\typeout{** loaded for the language `#1'. Using the pattern for}%
\typeout{** the default language instead.}%
\else
\language=\csname l@#1\endcsname
\fi
#2}}
\providecommand{\BIBdecl}{\relax}
\BIBdecl

\bibitem{AnalogSIC2}
J.~I. Choi, M.~Jain, K.~Srinivasan, P.~Levis, and S.~Katti, ``Achieving single
  channel, full duplex wireless communication,'' \emph{ACM MobiCom}, pp. 1--12,
  Sep. 2010.

\bibitem{bharadia2013full}
D.~Bharadia, E.~McMilin, and S.~Katti, ``Full duplex radios,'' \emph{SIGCOMM
  Comput. Commun. Rev.}, vol.~43, no.~4, pp. 375--386, Aug. 2013.

\bibitem{FDsurvey}
D.~Kim, H.~Lee, and D.~Hong, ``{A Survey of In-band Full-duplex Transmission:
  From the Perspective of PHY and MAC Layers},'' \emph{IEEE Commun. Surveys
  Tuts.}, no.~99, Feb. 2015.

\bibitem{distributedFD}
X.~Fang, D.~Yang, and G.~Xue, ``Distributed algorithms for multipath routing in
  full-duplex wireless networks,'' \emph{IEEE MASS}, pp. 102--111, Oct. 2011.

\bibitem{directionalFD}
K.~Kato and M.~Bandai, ``Routing protocol for directional full-duplex
  wireless,'' \emph{IEEE PIMRC}, pp. 3239--3243, Sep. 2013.

\bibitem{DAFD-MAC}
Y.~Sugiyama, K.~Tamaki, S.~Saruwatari, and T.~Watanabe, ``A wireless
  full-duplex and multi-hop network with collision avoidance using directional
  antennas,'' in \emph{International Conference on Mobile Computing and
  Ubiquitous Networking (ICMU)}, Jan 2014, pp. 38--43.

\bibitem{opt_pwr_rout}
D.~Ramirez and B.~Aazhang, ``{Optimal Routing and Power Allocation for Wireless
  Networks with Imperfect Full-Duplex Nodes},'' \emph{IEEE Trans. Wireless
  Commun.}, vol.~12, no.~9, pp. 4692--4704, Sep. 2013.

\bibitem{energyeffsurvey}
J.~Li, D.~Cordes, and J.~Zhang, ``Power-aware routing protocols in ad hoc
  wireless networks,'' \emph{IEEE Wireless Commun.}, vol.~12, no.~6, pp.
  69--81, Dec. 2005.

\bibitem{duarte2012experiment}
M.~Duarte, C.~Dick, and A.~Sabharwal, ``Experiment-driven characterization of
  full-duplex wireless systems,'' \emph{IEEE Trans. Wireless Commun.}, vol.~11,
  no.~12, pp. 4296--4307, May. 2012.

\bibitem{Power_Control_MAC}
E.-S. Jung and N.~H. Vaidya, ``{A Power Control MAC Protocol for Ad hoc
  Networks},'' \emph{ACM MobiCom}, pp. 36--47, Sep. 2002.

\bibitem{distributedNetworkMAC}
M.~Al-Kadri, A.~Aijaz, and A.~Nallanathan, ``{An Energy-Efficient Full-Duplex
  MAC Protocol for Distributed Wireless Networks},'' \emph{IEEE Wireless
  Commun. Lett.}, vol.~5, no.~1, pp. 44--47, Feb. 2016.

\bibitem{ieee80211}
``{IEEE Standard for Information technology--Telecommunications and information
  exchange between systems Local and metropolitan area networks--Specific
  requirements - Part 11: Wireless LAN Medium Access Control (MAC) and Physical
  Layer (PHY) Specifications},'' \emph{IEEE Std 802.11-2016}, pp. 1--3534, Dec
  2016.

\bibitem{AODV}
\BIBentryALTinterwordspacing
C.~Perkins, E.~Belding-Royer, and S.~Das, ``{Adhoc On-demand Distance Vector
  (AODV) Routing},'' Internet Engineering Task Force, RFC 3561, 2003. [Online].
  Available: \url{http://www.rfc-editor.org/rfc/rfc3561.txt}
\BIBentrySTDinterwordspacing

\bibitem{air_exp}
B.~Chen, Y.~Qiao, O.~Zhang, and K.~Srinivasan, ``Airexpress: Enabling seamless
  in-band wireless multi-hop transmission,'' in \emph{ACM MobiCom}, Sept. 2015,
  pp. 566--577.

\bibitem{bpass}
L.~Chen \emph{et~al.}, ``Bipass: Enabling end-to-end full duplex,'' in
  \emph{ACM MobiCom}, Oct. 2017.

\end{thebibliography}
%

\begin{IEEEbiographynophoto}{M. Omar Al-Kadri} (M'17) received the B.Eng. degree in Computer Engineering from IUST, Syria, in 2010, the M.Sc. degree (with distinctoin) in Networking and Data communication from Kingston University, UK, in 2013, and the Ph.D degree in Telecommunication engineering from King's College London, UK, in 2017. He is now a lecturer in networking and cyber security at Robert Gordon University, UK. His current research interests include security of wireless communications with application to healthcare,  full-duplex communications, HetNets, and MAC/routing protocols.
\end{IEEEbiographynophoto}

\begin{IEEEbiographynophoto}{Adnan Aijaz} (M'14--SM'18) received his Ph.D degree in Telecommunications Engineering from King's College London (KCL), UK, in  2014. After a post-doctoral year at KCL, he moved to Toshiba Research Europe Ltd. where he is currently  a Senior Research Engineer. His recent research interests include 802.11-based WLANs, 5G cellular networks, Industrial IoT, Tactile Internet, and full-duplex communications. His publications have been featured in internationally renowned conferences and journals.

Prior to joining KCL, he worked in cellular industry for nearly 2.5 years in the areas of network performance management, optimization, and quality assurance. He  holds a B.E. degree in Electrical (telecom) Engineering from National University of Sciences and Technology (NUST), Pakistan.
\end{IEEEbiographynophoto}

\begin{IEEEbiographynophoto}{Arumugam Nallanathan} (M'00--SM'05--F'17) 
is
Professor of Wireless Communications and Head
of the Communication Systems Research group in
the School of Electronic Engineering and Computer
Science at Queen Mary University of London since
September 2017. Previously, he held academic positions with the Department of
Informatics at King's College London, UK and the
Department of Electrical and Computer Engineering, National University of
Singapore. His research interests include
5G Wireless Networks, Internet of Things and Molecular Communications. He
is a co-recipient of the Best Paper Awards presented at the IEEE International
Conference on Communications 2016 and the IEEE Global Communications
Conference 2017. He is an IEEE Distinguished Lecturer. He has been selected
as a Web of Science Highly Cited Researcher in 2016.

He has been an editor for various IEEE journals and served as the chair and member of numerous IEEE conferences. 
He received the IEEE Communications Society SPCE outstanding service award 2012 and
IEEE Communications Society RCC outstanding service award 2014.

\end{IEEEbiographynophoto}



%








\end{document}